\documentclass[conference]{IEEEtran}

\usepackage[british]{babel}
\usepackage[T1]{fontenc}
\usepackage[utf8]{inputenc}
\usepackage{lmodern}
\usepackage[english]{isodate}

\usepackage[ruled,norelsize]{algorithm2e}

\usepackage{amsmath}
\usepackage{amssymb}
\usepackage{amsthm}
\usepackage{bm}

\usepackage{graphicx}
\usepackage{pgf, tikz}
\usepackage{pgfplots}
\pgfplotsset{compat=1.17}
\usepgfplotslibrary{statistics}
\usepackage{placeins}

\usepackage{subfig}
\usepackage{pgfplots}

\usepackage{nameref}
\usepackage[hidelinks, unicode]{hyperref}

\usepackage[style=ieee]{biblatex} 
\usepackage[autostyle=true]{csquotes}
\usepackage{listingsutf8}
\usepackage{multirow}
\usepackage{orcidlink}
\usepackage{seqsplit}
\usepackage{siunitx}
\usepackage{xcolor}

\definecolor{sign}{HTML}{b02a2d}
\definecolor{direction}{HTML}{007900}
\definecolor{regime}{HTML}{8c399e}
\definecolor{characteristic}{HTML}{1f5dc2}
\definecolor{mantissa}{HTML}{636363}
\definecolor{error}{HTML}{BD002A}
\definecolor{cellbg}{HTML}{EDEDED}

\definecolor{p-sign}{HTML}{FF5454}
\definecolor{p-regime}{HTML}{CC9966}
\definecolor{p-regime-term}{HTML}{996633}
\definecolor{p-exponent}{HTML}{0080FF}
\definecolor{p-fraction}{HTML}{000000}

\definecolor{posit}{HTML}{b02a2d}
\definecolor{bfloat16}{HTML}{007900}
\definecolor{takum}{HTML}{1f5dc2}
\definecolor{float}{HTML}{636363}

\makeatletter
\def\lst@makecaption{%
  \def\@captype{table}%
  \@makecaption
}
\makeatother

\begin{document}

\title{%
	Streamlining SIMD ISA Extensions with Takum Arithmetic: A Case Study on Intel AVX10.2
}
\author{%
	\IEEEauthorblockN{%
		Laslo Hunhold\,\orcidlink{0000-0001-8059-0298}}
	\IEEEauthorblockA{%
		\textit{Parallel and Distributed Systems Group}\\
		\textit{University of Cologne}\\
		Cologne, Germany\\
		\href{mailto:Laslo Hunhold <hunhold@uni-koeln.de>}
		{\texttt{hunhold@uni-koeln.de}}
	}
}

\maketitle

\begin{abstract}
Modern microprocessors extend their instruction set architecture (ISA) with Single Instruction, Multiple Data (SIMD) operations to improve performance. The Intel Advanced Vector Extensions (AVX) enhance the x86 ISA and are widely supported in Intel and AMD processors. The latest version, AVX10.2, places a strong emphasis on low-precision, non-standard floating-point formats, including bfloat16 and E4M3/E5M2 float8 (OCP 8-bit Floating Point, OFP8), primarily catering to deep learning applications rather than general-purpose arithmetic. However, as these formats remain within the IEEE 754 framework, they inherit its limitations, introducing inconsistencies and added complexity into the ISA.
\par
This paper examines the recently proposed tapered-precision takum floating-point format, which has been shown to offer significant advantages over IEEE 754 and its derivatives as a general-purpose number format. Using AVX10.2 as a case study, the paper explores the potential benefits of replacing the multitude of floating-point formats with takum as a uniform basis. The results indicate a more consistent instruction set, improving readability and flexibility while offering potential for 8- and 16-bit general-purpose SIMD arithmetic.
\end{abstract}


\section{Introduction}\label{sec:introduction}
\begin{figure}[bp]
	\begin{center}
		\begin{tikzpicture}
			\draw[<->] (0.0, 0.7) -- (0.4, 0.7) node[above,pos=.5] {sign};
			\draw[<->] (0.4, 0.7) -- (4.5, 0.7) node[above,pos=.5] 
			{exponent};
			\draw[<->] (4.5, 0.7) -- (8.0, 0.7) node[above,pos=.5] {fraction};
			
			\draw (0,  0  ) rectangle (0.4,0.5) node[pos=.5] 
			{\textcolor{sign}{S}};
			\draw (0.4,0  ) rectangle (0.8,0.5) node[pos=.5] 
			{\textcolor{direction}{D}};
			\draw (0.8,0  ) rectangle (2.0,0.5) node[pos=.5] 
			{\textcolor{regime}{R}};
			\draw (2.0,0  ) rectangle (4.5,0.5) node[pos=.5] 
			{\textcolor{characteristic}{C}};
			\draw (4.5,0  ) rectangle (8.0,0.5) node[pos=.5] 
			{\textcolor{mantissa}{F}};
			
			\draw[<->] (0.0, -0.2) -- (0.4, -0.2) node[below,pos=.5] {$1$};
			\draw[<->] (0.4, -0.2) -- (0.8, -0.2) node[below,pos=.5] {$1$};
			\draw[<->] (0.8, -0.2) -- (2.0, -0.2) node[below,pos=.5] {$3$};
			\draw[<->] (2.0, -0.2) -- (4.5, -0.2) node[below,pos=.5] {$r$};
			\draw[<->] (4.5, -0.2) -- (8.0, -0.2) node[below,pos=.5] {$p$};
		\end{tikzpicture}
	\end{center}
	\caption{
		The $n$-bit takum representation, comprising a sign bit $\textcolor{sign}{S}$, a direction bit $\textcolor{direction}{D}$, three regime bits $\textcolor{regime}{R}$, $r \in \{0,\dots,7\}$ characteristic bits $\textcolor{characteristic}{C}$, and $p \in \{n-12,\dots,n-5\}$ fraction bits $\textcolor{mantissa}{F}$. Since takums are invariant under zero-extension, bit strings with $n < 12$ are zero-extended to $12$ bits for decoding.
	}
	\label{fig:takum-encoding}
\end{figure}
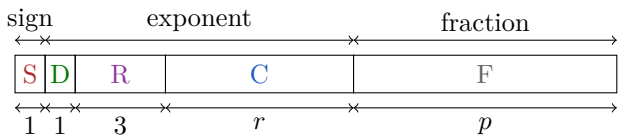
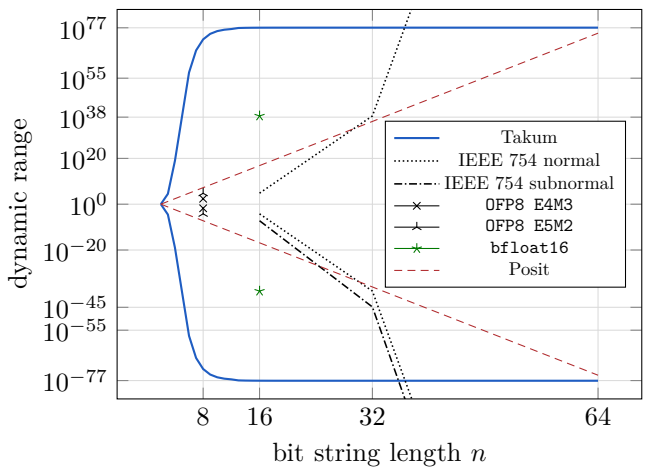
\begin{figure}[bp]
	\begin{center}
		\begin{tikzpicture}
			\begin{axis}[
				scale only axis,
				width=\textwidth/2.6,
				height=\textwidth/3.5,
				ymin=10^-85,
				ymax=10^85,
				xlabel={bit string length $n$},
				ylabel={dynamic range},
				ymode=log,
				ylabel shift=-0.1cm,
				xtick={8,16,32,64},
				xminorticks=true,
				yminorticks=true,
				ytick={10^-400,10^-300,10^-200,10^-100,10^-77,10^-55,10^-45,10^-20, 1,10^20,10^38,10^55,10^77,10^100,10^200,10^300,10^400},
				grid=both,
				minor y tick num=9,
				grid style={line width=.1pt, draw=gray!10},
				major grid style={line width=.2pt,draw=gray!30},
				legend style={nodes={scale=0.7, transform shape}},
				legend style={at={(0.97,0.5)},anchor=east}
			]
				\addplot[characteristic,thick] table [x=n, y=lintakum-min, col 
				sep=comma] {data/dynamic_range.csv};
				\addplot[characteristic,thick,forget plot] table [x=n, y=lintakum-max, col 
				sep=comma] {data/dynamic_range.csv};
				\addlegendentry{Takum};

				\addplot[semithick,densely dotted] table [x=n, y=ieee-normal-min, col sep=comma] {data/dynamic_range.csv};
				\addplot[forget plot,semithick,densely dotted] table [x=n, y=ieee-max, col sep=comma] {data/dynamic_range.csv};
				\addlegendentry{IEEE 754 normal};
				
				\addplot[semithick,densely dashdotted] table [x=n, y=ieee-subnormal-min, col sep=comma] {data/dynamic_range.csv};
				\addlegendentry{IEEE 754 subnormal};

				\addplot[mark=x,forget plot] coordinates {(8,0.016)};
				\addplot[mark=x] coordinates {(8,240)};
				\addlegendentry{\texttt{OFP8 E4M3}};

				\addplot[mark=Mercedes star,forget plot] coordinates {(8,6.103515625e-5)};
				\addplot[mark=Mercedes star] coordinates {(8,57344.0)};
				\addlegendentry{\texttt{OFP8 E5M2}};

				\addplot[direction,mark=star,forget plot] coordinates {(16,1.175494351e-38)};
				\addplot[direction,mark=star] coordinates {(16,3.38953139e38)};
				\addlegendentry{\texttt{bfloat16}};

				\addplot[sign,densely dashed] table [x=n, y=posit2-min, col sep=comma] {data/dynamic_range.csv};
				\addplot[sign,densely dashed,forget plot] table [x=n, y=posit2-max, col sep=comma] {data/dynamic_range.csv};
				\addlegendentry{Posit};
			\end{axis}
		\end{tikzpicture}
	\end{center}
	\caption{Dynamic range relative to the bit string length $n$ for linear takum, posit and a selection of floating-point formats. The bit-string lengths relevant to AVX10.2 are indicated on the x-axis}
	\label{fig:dynamic_range}
\end{figure}%
To maximise the utilisation of pipelined architectures in modern microprocessors, specialised instructions that operate on multiple data points within a single instruction---referred to as Single Instruction, Multiple Data (SIMD)---are integral to performance-oriented software. A prominent example of such an instruction set is Intel’s Advanced Vector Extensions (AVX) for x86. First introduced in 2008 and expanded as AVX2 in 2013, Intel proposed AVX-512 as a vector instruction set architecture (ISA) in the same year. Over the following years, AVX-512 was progressively extended (and fragmented) with more than 20 optional feature flags.
\par
This fragmentation, along with the desire for greater cross-platform consistency, motivated the development of a unified vector ISA known as AVX10 \cite{intel-avx10-paper}. The first version, AVX10.1, introduced in 2023, is functionally equivalent to AVX-512 and only explicitly enumerates AVX-512 instructions at 128, 256, and 512 bits \cite{intel-avx10.1}. The specification was further extended as AVX10.2 in 2024, most notably introducing support for 8-bit floating-point arithmetic \cite{intel-avx10.2} by implementing the OCP 8-Bit Floating Point Specification (OFP8) \cite{ofp8}, which defines the two formats E4M3 (four exponent bits, three significand bits) and E5M2 (five exponent bits, two significand bits). While, at the time of writing, no AMD processor supports AVX10, the relevant registers have already been reserved for future use \cite[Figure 3.4, registers \%xmm16--\%xmm31]{amd64-abi}.
\par%
Shortcomings in IEEE 754 floating-point arithmetic have led to the emergence of derivative number formats such as \texttt{bfloat16} \cite{bfloat16} and the aforementioned OFP8 types, as well as alternative formats such as posit arithmetic \cite{posits-beating_floating-point-2017} and, more recently, takum arithmetic \cite{2024-takum} (see Figure~\ref{fig:takum-encoding}). A key innovation in posits and takums, both of which are uniformly defined for arbitrary bit-string lengths, is their \enquote{tapered precision}. This property ensures a higher density of representable values near 1---where values are more frequently encountered in computations---at the expense of reduced density for values further from 1, which are used less frequently. Other advantages include a unique unsigned zero representation (in contrast to the signed zero in IEEE 754), a natural two’s complement ordering and negation, and simpler, more power-efficient hardware implementations \cite{posit-hardware-2018, posit-hardware_cost-2019, 2024-takum-hardware}.
\par
Takum arithmetic additionally features a consistently large dynamic range, which is nearly fully realised even at 8 bits, as illustrated in Figure~\ref{fig:dynamic_range}. This design choice reflects the observation that low-precision arithmetic typically involves a trade-off between precision and dynamic range, whereas ideally, only the former should be variable. Counterintuitively, takums demonstrate superior overall numerical performance despite their high dynamic range compared to IEEE 754 floating-point numbers and their derivatives \cite{2024-takum-benchmarks, 2025-takum-eigen, 2025-takum-fft}.
\par
From a hardware perspective, takum arithmetic exhibits the distinctive property that all precision levels share a common decoder, which is required to process at most the 12 most significant bits (see Figure~\ref{fig:takum-encoding}), as the remaining bits are guaranteed to be fraction bits. This stands in contrast to posits, where the encoded exponent may extend across the entire bit string. The takum design ensures highly consistent performance across varying precisions and surpasses that of state-of-the-art posit hardware codecs, which necessitate separate decoders for each precision~\cite{2024-takum-hardware}. A further notable advantage over posits is that 64-bit takums possess at least as many fraction bits as 64-bit IEEE 754 floating-point numbers, thereby achieving at least equivalent machine precision.
\par
This paper first examines the viability of takum arithmetic as a potential replacement for the diverse arithmetic formats present in AVX10.2. To this end, we devise a benchmark to assess the representational capabilities of these formats using a large, diverse set of sparse matrices across different precisions. Based on these findings, we then propose a set of guidelines for streamlining AVX10.2 and apply them to the instruction set.
\par
The remainder of this paper is structured as follows. Section~\ref{sec:viability} discusses the viability of takum arithmetic as a foundational arithmetic format for AVX10.2. Section~\ref{sec:methodology} details the methodology for grouping, modifying, and optimising the AVX10.2 instruction set. Section~\ref{sec:evaluation} presents an evaluation of the results, while Section~\ref{sec:conclusion} summarises our findings and provides concluding remarks.
\section{Viability of Takum Arithmetic}\label{sec:viability}
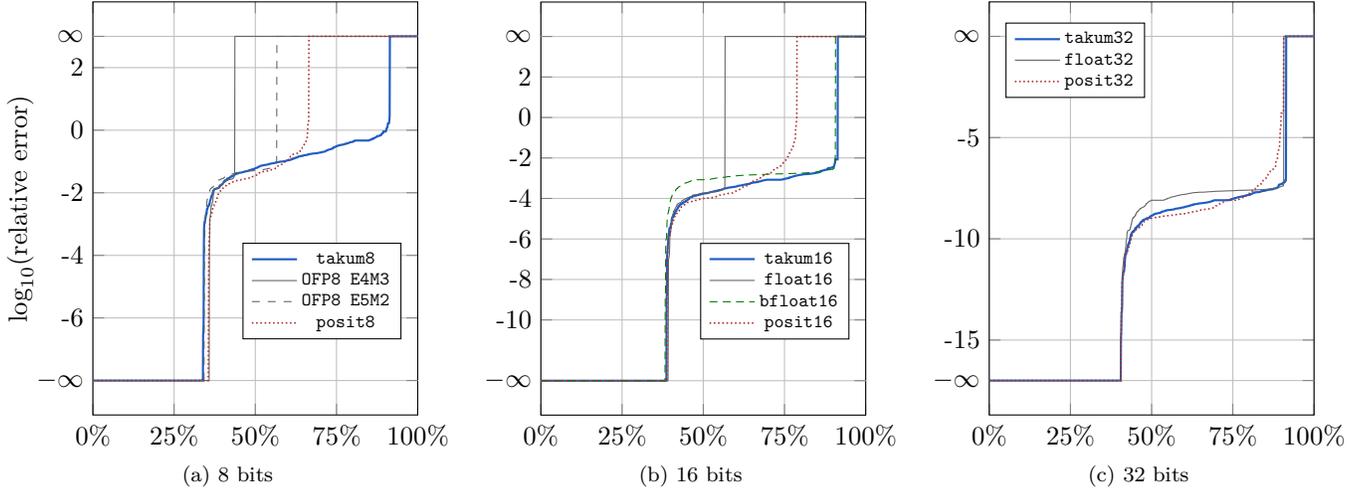
\begin{figure*}[tb]
	\begin{center}
        \subfloat[8 bits]{
    		\begin{tikzpicture}
    			\begin{axis}[
    				scaled y ticks={base 10:-2},
    				ytick={1e-6,1e-4,1e-2,1,1e2},
    				yticklabels={-6,-4,-2,0,2},
    				extra y ticks={1e-8,1e3},
    				extra y tick labels={$-\infty$,$\infty$},
    				grid=major,
    				ymode=log,
    				xmin=0,
    				xmax=1,
    				ylabel={$\log_{10}(\text{relative error})$},
                    ylabel shift=-0.2cm,
    				scale only axis,
    				width=\textwidth/4.2,
    				height=\textwidth/3.3,
    				xtick={0,0.25,0.5,0.75,1.0},
    				xticklabel={\pgfmathparse{\tick*100}\pgfmathprintnumber{\pgfmathresult}\%},
    				point meta={x*100},
    				legend style={nodes={scale=0.7, transform shape}},
    				legend style={at={(0.95,0.3)},anchor=east},
    				]
    				\addplot[takum,thick] table [col sep=comma, x=percent, y=LinearTakum8]{data/convert/relative_error.sorted.csv};
    				    \addlegendentry{\texttt{takum8}};
    				\addplot[float] table [col sep=comma, x=percent, y=Float8_4]{data/convert/relative_error.sorted.csv};
    				    \addlegendentry{\texttt{OFP8 E4M3}};
    				\addplot[float,dashed] table [col sep=comma, x=percent, y=Float8_5]{data/convert/relative_error.sorted.csv};
    				    \addlegendentry{\texttt{OFP8 E5M2}};
    				\addplot[posit,densely dotted,semithick] table [col sep=comma, x=percent, y=Posit8]{data/convert/relative_error.sorted.csv};
    				    \addlegendentry{\texttt{posit8}};
    			\end{axis}
    		\end{tikzpicture}
        }
        \subfloat[16 bits]{
    		\begin{tikzpicture}
    			\begin{axis}[
    				scaled y ticks={base 10:-2},
    				ytick={1e-10,1e-8,1e-6,1e-4,1e-2,1,1e2},
    				yticklabels={-10,-8,-6,-4,-2,0,2},
    				extra y ticks={1e-13,1e4},
    				extra y tick labels={$-\infty$,$\infty$},
    				grid=major,
    				ymode=log,
    				xmin=0,
    				xmax=1,
    				scale only axis,
    				width=\textwidth/4.2,
    				height=\textwidth/3.3,
    				xtick={0,0.25,0.5,0.75,1.0},
    				xticklabel={\pgfmathparse{\tick*100}\pgfmathprintnumber{\pgfmathresult}\%},
    				point meta={x*100},
    				legend style={nodes={scale=0.7, transform shape}},
    				legend style={at={(0.95,0.3)},anchor=east},
    				]
    				\addplot[takum,thick] table [col sep=comma, x=percent, y=LinearTakum16]{data/convert/relative_error.sorted.csv};
    				    \addlegendentry{\texttt{takum16}};
    				\addplot[float] table [col sep=comma, x=percent, y=Float16]{data/convert/relative_error.sorted.csv};
    				    \addlegendentry{\texttt{float16}};
    				\addplot[bfloat16,densely dashed] table [col sep=comma, x=percent, y=BFloat16]{data/convert/relative_error.sorted.csv};
    				    \addlegendentry{\texttt{bfloat16}};
    				\addplot[posit,densely dotted,semithick] table [col sep=comma, x=percent, y=Posit16]{data/convert/relative_error.sorted.csv};
    				    \addlegendentry{\texttt{posit16}};
    			\end{axis}
    		\end{tikzpicture}
        }
        \subfloat[32 bits]{
    		\begin{tikzpicture}
    			\begin{axis}[
    				scaled y ticks={base 10:-2},
    				ytick={1e-15,1e-10,1e-5},
    				yticklabels={-15,-10,-5},
    				extra y ticks={1e-17,1e0},
    				extra y tick labels={$-\infty$,$\infty$},
    				grid=major,
    				ymode=log,
    				xmin=0,
    				xmax=1,
    				scale only axis,
    				width=\textwidth/4.2,
    				height=\textwidth/3.3,
    				xtick={0,0.25,0.5,0.75,1.0},
    				xticklabel={\pgfmathparse{\tick*100}\pgfmathprintnumber{\pgfmathresult}\%},
    				point meta={x*100},
    				legend style={nodes={scale=0.7, transform shape}},
    				legend style={at={(0.05,0.95)},anchor=north west},
    				]
    				\addplot[takum,thick] table [col sep=comma, x=percent, y=LinearTakum32]{data/convert/relative_error.sorted.csv};
    				    \addlegendentry{\texttt{takum32}};
    				\addplot[float] table [col sep=comma, x=percent, y=Float32]{data/convert/relative_error.sorted.csv};
    				    \addlegendentry{\texttt{float32}};
    				\addplot[posit,densely dotted,semithick] table [col sep=comma, x=percent, y=Posit32]{data/convert/relative_error.sorted.csv};
    				    \addlegendentry{\texttt{posit32}};
    			\end{axis}
    		\end{tikzpicture}
        }
	\end{center}
	\caption{
        Cumulative error distribution of the relative 2-norm errors of
        the matrices after conversion to a range of machine number types.
        The symbol $\infty$ denotes where the dynamic range of the
        matrix entries exceeded the target number type.
	}
	\label{fig:conversion_errors}
\end{figure*}
The variety of formats available in AVX10.2 (\texttt{bfloat16}, OFP8 E4M3, OFP8 E5M2, alongside standard formats) reflects the absence of a general-purpose low-precision format in the current ecosystem. Ultimately, E4M3 and E5M2 at 8 bits, as well as \texttt{float16} and \texttt{bfloat16} at 16 bits, represent different parameterisations that balance precision and dynamic range. As discussed earlier, takum arithmetic may serve as a viable alternative, as its tapered precision enables a high dynamic range while maintaining high precision for values near 1. This section empirically evaluates this hypothesis through a representative benchmark.
\par
Although previous studies have examined takum arithmetic from various perspectives \cite{2024-takum-benchmarks,2024-takum-hardware,2024-takum-integers}, the benchmark in this section focuses specifically on the representational capabilities of different formats. To construct a comprehensive dataset, we select all matrices from the SuiteSparse Matrix Collection \cite{davis2011university} with at most 50,000 non-zero entries, resulting in a total of 1,401 matrices. The dataset is prepared using the MuFoLAB (Multi-Format Linear Algebra Benchmarks) tool \cite{mufolab}, as introduced in \cite{2024-takum-benchmarks}. The SuiteSparse Matrix Collection encompasses matrices from a diverse range of real-world applications, including (but not limited to) computational fluid dynamics, chemical simulation, materials science, optimal control, structural mechanics, and 2D/3D sequencing. The benchmark procedure consists of converting each matrix into the formats under consideration, followed by conversion to \texttt{float128} (using \texttt{Quadmath.jl}) to compute the relative 2-norm error \cite[src/convert.jl]{mufolab}.
\par
The results, presented in Figure~\ref{fig:conversion_errors}, highlight clear differences in accuracy. At 8 bits, neither OFP8 E4M3 nor OFP8 E5M2 exhibit satisfactory stability, with relative errors exceeding \SI{100}{\percent} for approximately \SI{45}{\percent} and \SI{55}{\percent} of test matrices, respectively. Posits outperform both formats in terms of precision and stability, with approximately \SI{65}{\percent} of matrices remaining below \SI{100}{\percent} relative error. While takums provide slightly lower precision than posits initially, they demonstrate superior overall stability, with around \SI{90}{\percent} of matrices below \SI{100}{\percent} relative error, and comparable performance to OFP8 formats within their respective stability regions.
\par
At 16 bits, takum arithmetic consistently surpasses both \texttt{float16} and \texttt{bfloat16}, whereas posits exhibit lower overall stability despite achieving lower errors initially. This trend persists at 32 bits, where takums outperform \texttt{float32} across the board, while posits display a region with higher errors than \texttt{float32}, despite demonstrating marginally better performance at lower errors.
\par
Although further analysis is warranted, the benchmark results indicate that takum arithmetic is a robust general-purpose numerical format, outperforming all arithmetic formats in AVX10.2 as well as posit arithmetic. Our results demonstrate that posit arithmetic does not consistently surpass \texttt{bfloat16} and \texttt{float32}, whereas takum arithmetic exhibits clear advantages across multiple precision levels. These findings provide strong evidence for takum's viability for AVX10.2.
\section{Methodology}\label{sec:methodology}
In this section, we outline potential approaches for reformulating the AVX10.2 instruction set. These methods are derived from observations of the instruction set, as detailed in Tables~\ref{tab:bitwise_instructions}--\ref{tab:cryptographic_instructions}, which we reference here in advance for conciseness. In addition to transitioning to takum arithmetic, we also propose further modifications to enhance the instruction set.
\begin{enumerate}
	\item{
		\textbf{Instruction Grouping}: Although not explicitly defined by the ISA, we classify instructions into the following groups: bitwise, mask, integer, floating-point, and cryptographic. Conversion routines involving floating-point numbers in any capacity are assigned to the floating-point category. This classification provides a clearer overview of the instruction set and its organisation.
	}
	\item{
		\textbf{Bit Quantity Naming}: Within the instruction set, bit quantities are conventionally referred to as bytes (8 bits, denoted by \texttt{B}), words (16 bits, \texttt{W}), double words (32 bits, \texttt{D}), and quad words (64 bits, \texttt{Q}). While these conventions remain widely used, they do not scale effectively beyond 64 bits, particularly given the inconsistent definitions of the term \enquote{word} across different architectures. Additionally, the extensive use of abbreviations in instruction names reduces readability. Furthermore, operations on raw bits and integer values are often indistinguishable in the current naming scheme, while signed and unsigned integers are denoted separately using the letters \texttt{S} and \texttt{U}.   
		\par
		To improve clarity, scalability, and consistency, we propose adopting a systematic notation: \texttt{B8}, \texttt{B16}, \texttt{B32}, \texttt{B64}, etc., for bitwise interpretation of singular values or packed vectors of 8, 16, 32, and 64 bits. Similarly, unsigned integers are represented as \texttt{U8}, \texttt{U16}, \texttt{U32}, \texttt{U64}, etc., while signed integers are denoted as \texttt{S8}, \texttt{S16}, \texttt{S32}, \texttt{S64}, etc.
	}
	\item{
		\textbf{Floating-Point Naming}: Analogous to bit quantity naming, floating-point formats in the instruction set are referred to using legacy terminology: \texttt{float16} is designated as half-precision (\texttt{H}), \texttt{float32} as single-precision (\texttt{S}), and \texttt{float64} as double-precision (\texttt{D}). However, this convention is unsustainable in both directions. For instance, \texttt{float128} (quad precision) cannot use \texttt{Q}, as this designation is already assigned to quad words, and similarly, quarter-precision 8-bit floating-point numbers cannot be assigned \texttt{Q} either.
		\par
		Furthermore, non-standard arithmetic formats introduce additional complexity. For example, OFP8 E4M3 is referred to as \texttt{HF8}, OFP8 E5M2 as \texttt{BF8}, and \texttt{bfloat16} as \texttt{BF16}. Complexity is further compounded by special biased operations for OFP8 types (e.g., \texttt{VCVTBIASPH2BF8}) and exception-free operations specific to \texttt{bfloat16} (e.g., \texttt{VDIVNEPBF16}).
		\par
		In this work, we simplify the nomenclature by omitting special number types and instead refer to 8-, 16-, 32-, and 64-bit packed or single takums as \texttt{T8}, \texttt{T16}, \texttt{T32}, and \texttt{T64}, respectively.
	}
	\item{
		\textbf{Generalisation}: To ensure consistency, we extend instructions that are currently restricted to specific precisions to cover the full range. This approach is justified by the removal of numerous instructions associated with non-standard floating-point formats, as well as the fact that all takum precisions share a common decoder.
	}
\end{enumerate}
Apart from this general approach, additional minor modifications will be applied depending on the specific instructions under consideration.
\section{Evaluation}\label{sec:evaluation}
AVX10.2 comprises 756 instructions, categorised as follows: 220 bitwise instructions (Table~\ref{tab:bitwise_instructions}), 59 mask instructions (Table~\ref{tab:mask_instructions}), 107 integer instructions (Table~\ref{tab:integer_instructions}), 363 floating-point instructions (Table~\ref{tab:floating_point_instructions}), and 7 cryptographic instructions (Table~\ref{tab:cryptographic_instructions}). To concisely represent these instructions, we employ standard regular expression notation in the tables. The following sections examine each category and analyse the modifications resulting from the application of the methods outlined in Section~\ref{sec:methodology}.
\subsection{Bitwise Instructions}
\begin{table}[tbp]
	\caption{
		AVX10.2 bitwise instructions, grouped together with regular expressions and
		assigned group IDs, and their respective proposed instructions.
	}
	\label{tab:bitwise_instructions}
	\centering
	\bgroup
	\def\arraystretch{1.2}
	\setlength{\tabcolsep}{0.5em}
	\begin{tabular}{| p{0.5cm} | p{3.7cm} | p{3.7cm} |}
		\hline
		\centering \textbf{ID} & \textbf{AVX10.2 instructions} & \textbf{proposed instructions}\\\hline
		\hline
		\centering B01 &
			\texttt{\seqsplit{V(ALIGN|PCONFLICT|P(GATHER|SCATTER)(D|Q)|PLZCNT|PRO(L|R)V?|PTERNLOG)(D|Q)}} &
			\multirow{3}{3.7cm}{\texttt{\seqsplit{V(ALIGN|ANDN?P|BLENDMP|COMPRESSP|CVTUSI2S|EXPANDP|EXTR|(GATHER|SCATTER)B(32|64)P|INSR|MOV(NT)?P(S|D)|PBLENDM|PCOMPRESS|PCONFLICT|PERM(I2|T2)?|PERM(IL|I2|T2)?P|PEXPAND|P(GATHER|SCATTER)B(32|64)|PLZCNT|PRO(L|R)V?|PTERNLOG|PTESTN?M|RANGE(P|S)|SHUFP|UNPCL(L|H)P|X?ORP)B(8|16|32|64)}}}
			\\\cline{1-2}
		\centering B02 &
			\texttt{\seqsplit{V(ANDN?P|BLENDMP|COMPRESSP|CVTUSI2S|EXPANDP|EXTR|(GATHER|SCATTER)(D|Q)P|INSR|PBLENDM|PCOMPRESS|PERM(I2|T2)?|PERM(IL|I2|T2)?P|PEXPAND|PTESTN?M|RANGE(P|S)|SHUFP|UNPCL(L|H)P|X?ORP)(S|D)}} &
			\\\cline{1-2}
		\centering B03 &
			\texttt{\seqsplit{VMOV((D|S(L|H))DUP|(LH|HL)PS|(L|H|A|U|NT)P(S|D)|S(H|S|D))|VMOV(D(Q(A(32|64)?|U(8|16|32|64)?))?|NTDQA?|Q|W)}} &
			\\\hline
		\centering B04 &
			\texttt{\seqsplit{VBROADCAST(F(32X(2|4|8)|64X(2|4)|128)|S(S|D))}} &
			\multirow{4}{3.7cm}{\texttt{\seqsplit{V(BROADCAST|EXTRACT|INSERT|P?SHUF|PS[L,R]L|PSRA|PUNPCK(H|L))B(8|16|32|64|128|256)}}}
			\\\cline{1-2}
		\centering B05 &
			\texttt{\seqsplit{VPBROADCAST(B|W|D|Q|M)?}} &
			\\\cline{1-2}
		\centering B06 &
			\texttt{\seqsplit{V(EXTRACT|INSERT)((F|I)(32X4|32X8|64X2|64X4|128)|PS)}} &
			\\\cline{1-2}
		\centering B07 &
			\texttt{\seqsplit{VSHUF[F,I][32x4,64x2]}} &
			\texttt{\seqsplit{}}
			\\\cline{1-2}
		\centering B08 &
			\texttt{\seqsplit{VPSHUF(B|HW|LW|D|BITQMB)}} &
			\texttt{\seqsplit{}}
			\\\cline{1-2}
		\centering B09 &
			\texttt{\seqsplit{VPS[L,R]L[D,DQ,Q,VD,VQ,VW,W]}} &
			\texttt{\seqsplit{}}
			\\\cline{1-2}
		\centering B10 &
			\texttt{\seqsplit{VPSRA[D,Q,VD,VQ,VW,W]}} &
			\texttt{\seqsplit{}}
			\\\cline{1-2}
		\centering B11 &
			\texttt{\seqsplit{VPUNPCK[H,L][BW,WD,DQ,QDQ]}} &
			\texttt{\seqsplit{}}
			\\\hline
		\centering B12 &
			\texttt{\seqsplit{VP(ALIGNR|ANDN?|MULTISHIFTQB|OPCOUNT|SH(L|R)DV?|X?OR)}} &
			\texttt{\seqsplit{VP(ALIGNR|ANDN?|MULTISHIFTQB|OPCOUNT|SH(L|R)DV?|X?OR)}}
			\\\hline
	\end{tabular}
	\egroup
\end{table}
The bitwise instruction set consists of 12 distinct groups, as shown in Table~\ref{tab:bitwise_instructions}. By applying the proposed modifications, groups B01--B03 can be unified into a single group covering operations on 8- to 64-bit values, while groups B04--B11 can be merged into one covering 8- to 256-bit operations. The B12 group remains unchanged. 
\par
A key consideration is the distinction between bitwise operations on floating-point numbers and integers. While floating-point and integer registers are technically separate, takums can be processed in a manner similar to signed integers for comparison operations. Consequently, this distinction may no longer be necessary. More broadly, it should be examined whether such a separation is justified at the instruction set level.
\subsection{Mask Instructions}
\begin{table}[tbp]
	\caption{
		AVX10.2 mask instructions, grouped together with regular expressions and
		assigned group IDs, and their respective proposed instructions.
	}
	\label{tab:mask_instructions}
	\centering
	\bgroup
	\def\arraystretch{1.2}
	\setlength{\tabcolsep}{0.5em}
	\begin{tabular}{| p{0.5cm} | p{3.7cm} | p{3.7cm} |}
		\hline
		\centering \textbf{ID} & \textbf{AVX10.2 instructions} & \textbf{proposed instructions}\\\hline
		\hline
		\centering M01 &
			\texttt{\seqsplit{K(ADD|ANDN?|MOV|NOT|OR(TEST)?|SHIFTL|SHIFTR|TEST|XNOR|OR)(B|W|D|Q)}} &
			\texttt{\seqsplit{K(ADD|ANDN?|MOV|NOT|OR(TEST)?|SHIFTL|SHIFTR|TEST|XNOR|OR)B(8|16|32|64)}}
			\\\hline
		\centering M02 &
			\texttt{\seqsplit{VKUNPCK(BW|WD|DQ)}} &
			\texttt{\seqsplit{VKUNPCK(B8B16|B16B32|B32B64)}}
			\\\hline
		\centering M03 &
			\texttt{\seqsplit{VPMOV(B|W|D|Q)2M}} &
			\texttt{\seqsplit{VPMOVB(8|16|32|64)2M}}
			\\\hline
		\centering M04 &
			\texttt{\seqsplit{VPMOVM2(B|W|D|Q)}} &
			\texttt{\seqsplit{VPMOVM2B(8|16|32|64)}}
			\\\hline
	\end{tabular}
	\egroup
\end{table}
As shown in Table~\ref{tab:mask_instructions}, the modifications to mask instructions are minimal, consisting primarily of adjustments related to the revised bit quantity naming convention.
\subsection{Integer Instructions}
\begin{table}[tbp]
	\caption{
		AVX10.2 integer instructions, grouped together with regular expressions and
		assigned group IDs, and their respective proposed instructions.
	}
	\label{tab:integer_instructions}
	\centering
	\bgroup
	\def\arraystretch{1.2}
	\setlength{\tabcolsep}{0.5em}
	\begin{tabular}{| p{0.5cm} | p{3.7cm} | p{3.7cm} |}
		\hline
		\centering \textbf{ID} & \textbf{AVX10.2 instructions} & \textbf{proposed instructions}\\\hline
		\hline
		\centering I01 &
			\texttt{\seqsplit{V(DBP|MP|P)SADBW}} &
			\texttt{\seqsplit{V(DBP|MP|P)SADU8U16}}
			\\\hline
		\centering I02 &
			\texttt{\seqsplit{VP(ABS|ADD|CMP|CMPEQ|CMPGT|CMPU|MAX(S|U)|MIN(S|U)|SUB)(B|W|D|Q)}} &
			\multirow{2}{3.7cm}{\texttt{\seqsplit{VP(ABSS|ADDU|CMPS|CMPEQU|CMPGTS|CMPUS|MAX(S|U)|MIN(S|U)|SUBU)(8|16|32|64)}}}
			\\\cline{1-2}
		\centering I03 &
			\texttt{\seqsplit{VP(ADDU?S|AVG|SUBU?S)(B|W)}} &
			\\\hline
		\centering I04 &
			\texttt{\seqsplit{VPACK(S|U)S(DW|WB)}} &
			\texttt{\seqsplit{VPACK(S|U)(S32S16|S16S8)}}
			\\\hline
		\centering I05 &
			\texttt{\seqsplit{VPCLMULQDQ}} &
			\texttt{\seqsplit{VPCLMULS64}}
			\\\hline
		\centering I06 &
			\texttt{\seqsplit{VPDP(B|W)(S|U)(S|U)DS?}} &
			\texttt{\seqsplit{VPDP(U8|U16)(S|U)(S|U)DS?}}
			\\\hline
		\centering I07 &
			\texttt{\seqsplit{VPMADD(52(L|H)UQ|UBSW|WD)}} &
			\texttt{\seqsplit{VPMADD(52(L|H)U64|U8S16|S16S32)}}
			\\\hline
		\centering I08 &
			\texttt{\seqsplit{VPMOV(WB|DB|DW|QB|QW|QD|(S|Z)X)}} &
			\texttt{\seqsplit{VPMOV(S16S8|S32S8|S32S16|S64S8|S64S16|S64S32)}}
			\\\hline
		\centering I09 &
			\texttt{\seqsplit{VPMUL(DQ|H(RS|U)?W|L(W|D|Q)|UDQ)}} &
			\texttt{\seqsplit{VPMUL(L|H)?U(8|16|32|64)}}
			\\\hline
	\end{tabular}
	\egroup
\end{table}
Regarding the integer instructions, listed with their respective modifications in Table~\ref{tab:integer_instructions}, a key change is that the instructions in groups I02 and I03 have been amended to consistently reflect the signedness of the integers they operate on. Other modifications primarily involve renaming instructions to align with the revised bit quantity naming scheme.
\subsection{Floating-Point Instructions}
\begin{table}[tbp]
	\caption{
		AVX10.2 floating-point instructions, grouped together with regular expressions and
		assigned group IDs, and their respective proposed instructions.
	}
	\label{tab:floating_point_instructions}
	\centering
	\bgroup
	\def\arraystretch{1.2}
	\setlength{\tabcolsep}{0.5em}
	\begin{tabular}{| p{0.5cm} | p{3.7cm} | p{3.7cm} |}
		\hline
		\centering \textbf{ID} & \textbf{AVX10.2 instructions} & \textbf{proposed instructions}\\\hline
		\hline
		\centering F01 &
			\texttt{\seqsplit{V(ADD|FN?M(ADD|SUB)(132|213|231)|MINMAX|MUL|REDUCE|RNDSCALE|SQRT|SUB)(NEPBF16|(P|S)(H|S|D))}} &
			\multirow{6}{3.7cm}{\texttt{\seqsplit{V(ADD|CLASS|DIV|EXP|FC?(MADD|MUL)C|FIXUPIMM|FM(ADDSUB|SUBADD)(132|213|231)|FN?M(ADD|SUB)(132|213|231)|MANT|MAX|MIN|MINMAX|MUL|RANGE|R(CP|SQRT)|REDUCE|RNDSCALE|SCALE|SQRT|SUB|U?CMP)(P|S)T(8|16|32|64)}}}
			\\\cline{1-2}
		\centering F02 &
			\texttt{\seqsplit{V(FIXUPIMM|RANGE)(P|S)(S|D)}} &
			\\\cline{1-2}
		\centering F03 &
			\texttt{\seqsplit{V(CMP|FPCLASS|GET(EXP|MANT)|MIN|MAX|SCALEF)(PBF16|(P|S)(H|S|D))|VCOMSBF16}} &
			\\\cline{1-2}
		\centering F04 &
			\texttt{\seqsplit{V(U?COM(I|X)S|DIV(P|S)|FM(ADDSUB|SUBADD)(132|213|231)P)(H|S|D)|VDIVNEPBF16}} &
			\\\cline{1-2}
		\centering F05 &
			\texttt{\seqsplit{VF(C?(MADD|MUL))C(P|S)H}} &
			\\\cline{1-2}
		\centering F06 &
			\texttt{\seqsplit{VR(CP|SQRT)(14(P|S)(S|D)|P(BF16|H)|SH)}} &
			\\\hline
		\centering F07 &
			\texttt{\seqsplit{VCVT2PS2PHX|VCVT(BIAS|NE2?)PH2(B|H)F8S?|VCVTHF82PH|VCVTNE2?PS2BF16|VCVTNEBF162IU?BS|VCVTPD2(DQ|P(H|S)|QQ|U(D|Q)Q)|VCVTPH2(DQ|IU?BS|P(S|SX|D)|QQ|U(D|Q)Q,U?W)|VCVTPS2(DQ|IU?BS|P(D|HX?)|QQ|U(D|Q)Q)|VCVTSD2S(H|S|I)|VCVTSH2(S(S|D|I)|USI)|VCVTSS2(S(H|D|I)|USI)|VCVTTNEBF162IU?BS|VCVTTPD2U?(D|Q)QS?|VCVTTPH2(DQ|IU?BS|U?(D|Q)Q|UW|W)|VCVTTPS2(IU?BS|U?(D|Q)QS?)|VCVTTS(D|S)2U?SIS?|VCVTTSH2U?SI|VCVTU?W2PH|VCVT(U?(D|Q)Q2P|SI2S)(H|S|D)|VCVT(U?(D|Q)Q2P|SI2S)(H|S|D)}} &
			\texttt{\seqsplit{VCVT(P(S|U)(8|16|32|64)2PT(8|16|32|64)|S(S|U)(8|16|32|64)2ST(8|16|32|64)|PT(8|16|32|64)2P(S|U)(8|16|32|64)|ST(8|16|32|64)2S(S|U)(8|16|32|64))}}
			\\\hline
		\centering F08 &
			\texttt{\seqsplit{VDPBF16PS|VDPPHPS}} &
			\texttt{\seqsplit{VDPPT8PT16|VDPPT16PT32|VDPPT32PT64}}
			\\\hline
	\end{tabular}
	\egroup
\end{table}
The transition to takum arithmetic and the removal of non-standard arithmetic formats, as detailed in Table~\ref{tab:floating_point_instructions}, result in a significant simplification of the instruction set. Specifically, groups F01--F06 can now be unified into a single group. Additionally, minor refinements have been introduced, such as the removal of \texttt{GET} from \texttt{VGET(EXP|MANT)(PBF16|(P|S)(H|S|D))} and \texttt{FP} from \texttt{VFPCLASS(PBF16|(P|S)(H|S|D))}.
\par
Many instructions have been extended to support additional precisions. For instance, groups F01, F03, and F04, which were previously limited to half-, single-, and double-precision operations, have been expanded to also accommodate 8-bit takums. Similarly, groups F02 and F06 now support both 8-bit and 16-bit takums.
\par
A major aspect of the modifications involves the overhaul of the conversion instructions in group F07. Due to the proliferation of arithmetic formats, the original instruction set suffered from considerable complexity and numerous special cases. The revised approach introduces a more streamlined and consistent conversion instruction set, eliminating redundant operations. Notably, certain operations, such as biased 8-bit conversions, have been removed, as they are unnecessary in takum arithmetic.
\subsection{Cryptographic Instructions}
\begin{table}[tbp]
	\caption{
		AVX10.2 cryptographic instructions, grouped together with regular expressions and
		assigned group IDs, and their respective proposed instructions.
	}
	\label{tab:cryptographic_instructions}
	\centering
	\bgroup
	\def\arraystretch{1.2}
	\setlength{\tabcolsep}{0.5em}
	\begin{tabular}{| p{0.5cm} | p{3.7cm} | p{3.7cm} |}
		\hline
		\centering \textbf{ID} & \textbf{AVX10.2 instructions} & \textbf{proposed instructions}\\\hline
		\hline
		\centering C01 &
			\texttt{\seqsplit{VAES(DEC|ENC)(LAST)?}} &
			\texttt{\seqsplit{VAES(DEC|ENC)(LAST)?}}
			\\\hline
		\centering C02 &
			\texttt{\seqsplit{VGF2P8AFFINE(INV)?QB}} &
			\texttt{\seqsplit{VGF2P8AFFINE(INV)?U64U8}}
			\\\hline
		\centering C03 &
			\texttt{\seqsplit{VGF2P8MULB}} &
			\texttt{\seqsplit{VGF2P8MULU8}}
			\\\hline
	\end{tabular}
	\egroup
\end{table}
The cryptographic instructions, similar to the mask instructions, have undergone only minor modifications to accommodate the updated bit quantity naming convention. These changes are summarised in Table~\ref{tab:cryptographic_instructions}.
\section{Conclusion}\label{sec:conclusion}
In this work, we explored takum arithmetic as a replacement for the diverse set of arithmetic formats present in the AVX10.2 vector ISA. We first demonstrated its viability through benchmarks on a large real-world dataset and subsequently outlined methods to restructure the AVX10.2 instruction set with a set of possible improvements. Our analysis revealed a notable simplification and greater consistency across the instruction set. Furthermore, the revised quantity naming convention enhances the extendability of the instruction set, ensuring greater adaptability for future architectures.
\par
This study is particularly significant as it presents a potential paradigm shift in computer arithmetic. Given further research and evaluation, it suggests a future where both low- and high-precision computations could be unified under a single number format, thereby substantially simplifying hardware design. While such a transition may seem radical, our findings, like many others, highlight a key insight: IEEE 754 floating-point numbers are poorly suited for modern low-precision computations. Despite numerous parametric extensions introduced at different precisions, they remain outperformed by a next-generation number format designed with efficiency and versatility in mind. Particular emphasis can be placed on 8- and 16-bit types, as they are crucial given that the memory wall imposes significant penalties on communication overheads.
\par
Future research can build upon our benchmarks by extending them to a broader range of datasets for increased coverage. Additionally, a more in-depth analysis of the instruction set could examine the interplay of floating-point exceptions and their implications for hardware implementation. Another promising direction is the study of corresponding RISC-V and ARM vector extensions, which have recently been proposed, to assess the broader applicability of takum arithmetic in diverse architectures.
\printbibliography
\end{document}